\documentclass[12pt,final]{article}
\usepackage{showkeys}\usepackage[dvips]{graphicx}
\usepackage[margin=3cm]{geometry}

\begin{document}

\title{\Large \bf Formation of capillary bridges in AFM-like geometry}\author{F. Dutka and M. Napi\'orkowski\\ 
Instytut Fizyki Teoretycznej, Uniwersytet Warszawski,\\  00-681 Warszawa, Ho\.za 69, Poland}
\maketitle{}
\abstract{We discuss the phase diagram of fluid confined in AFM-like geometry. It combines the 
properties of capillary condensation and complete filling of a wedge.}
\newpage
The properties of confined fluids are different from those in the bulk. 
A well known  example is  the phenomenon of 
capillary condensation and the corresponding phase diagram which displays the Kelvin law \cite{Widom,kelvinlaw}. 
In capillary of thickness $L$,  the liquid-vapor coexistence line $\mu = \mu_{L}(T)$ is shifted with respect to 
its bulk position $\mu = \mu_{0}(T)$ proportionally to $1/L$, i.e., $\mu_{0}(T)-\mu_{L}(T) \sim 1/L$, where 
$\mu$ denotes the chemical potential, and $T$ is the temperature.  Another example is a slit formed by two parallel 
and infinite walls, each of which preferentially adsorbs a different phase of the fluid \cite{ParryEvans,Binder}. 
In such a slit one observes a wetting dominated  phase transition in which - upon increasing the temperature - the 
position of the interface separating phases adsorbed at the walls settles in the center of the slit.  In case of a 
fluid confined in a wedge with the opening angle $\varphi$ one may observe - at bulk coexistence - the critical filling 
transition \cite{Rejmer,Hauge} in which the amount of liquid adsorbed in the wedge  grows infinitely when the 
temperature approaches the filling temperature $T_{f}(\varphi)$ such that the corresponding contact angle 
$\theta(T)$ takes the value $\theta(T_{f}(\varphi)) = \pi/2 - \varphi/2$. Note that $T_{f}(\pi) = T_{w}$, 
where $T_{w}$ is the wetting temperature of a planar substrate; in general $T_{f}(\varphi) \le T_{w}$. Also in a 
wedge  - but off bulk coexistence - one observes the complete filling transition in which the volume of the 
adsorbed liquid grows infinitely  upon approaching - at fixed temperature - the bulk coexistence from the vapor 
side.\\ 

\begin{figure}[htb]
\begin{center} 
\includegraphics{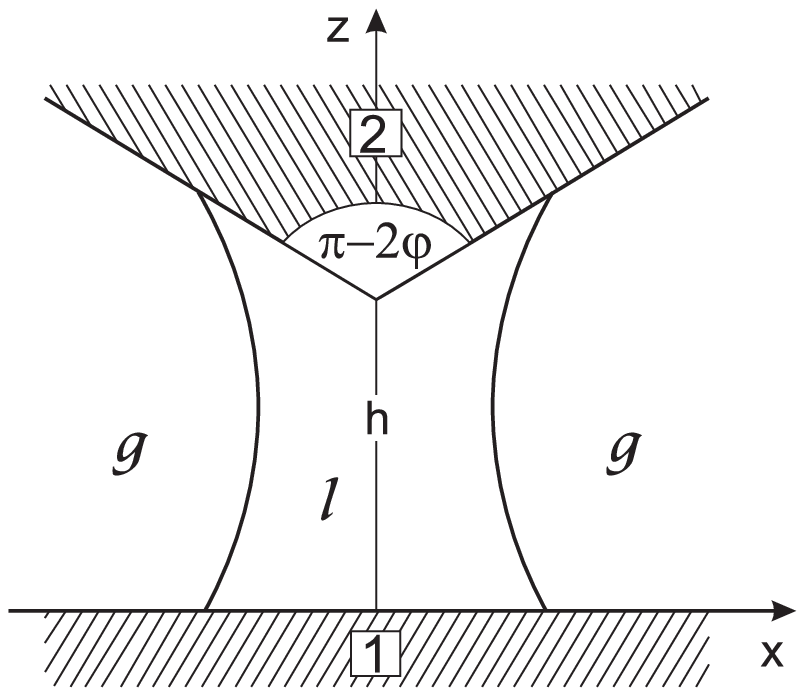} 
\caption{} 
\label{Fig. 1}  
\end{center} 
\end{figure}

\begin{figure}[htb] 
\begin{center}  
\includegraphics{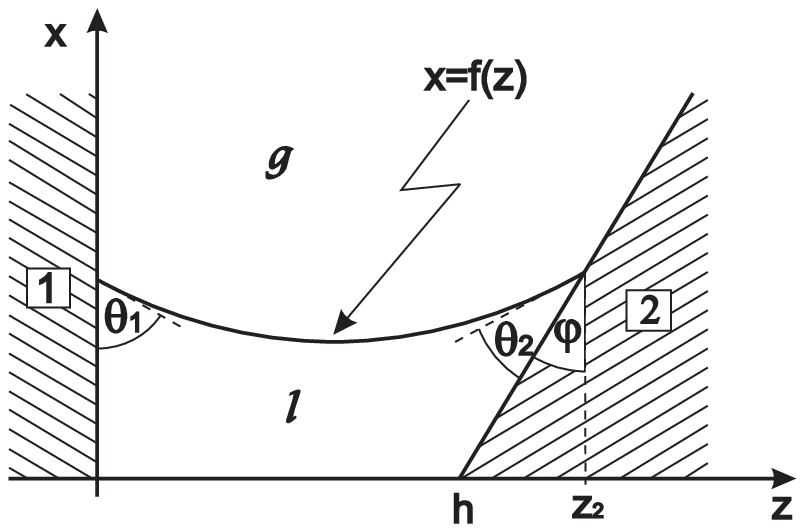} 
\caption{}
\label{Fig. 2}
\end{center}
\end{figure}

\begin{figure}[htb]
\begin{center}  
\includegraphics{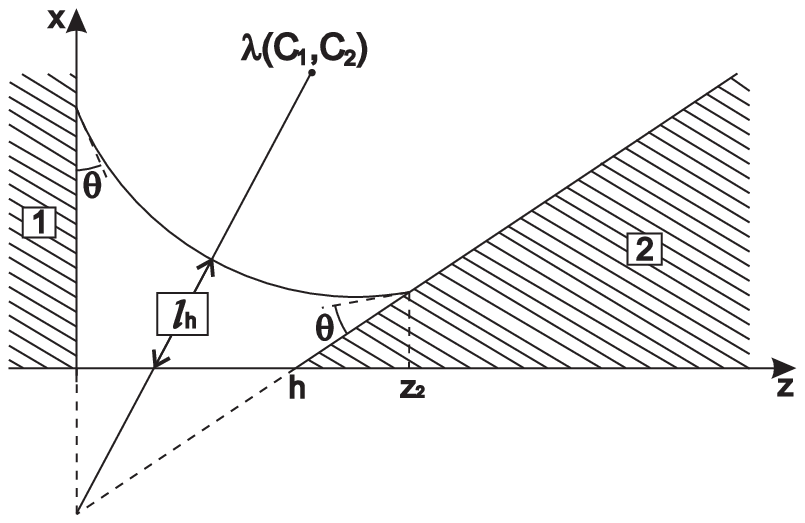} 
\caption{} 
\label{Fig. 3} 
\end{center} 
\end{figure}

\begin{figure}[htb] 
\begin{center}  
\includegraphics{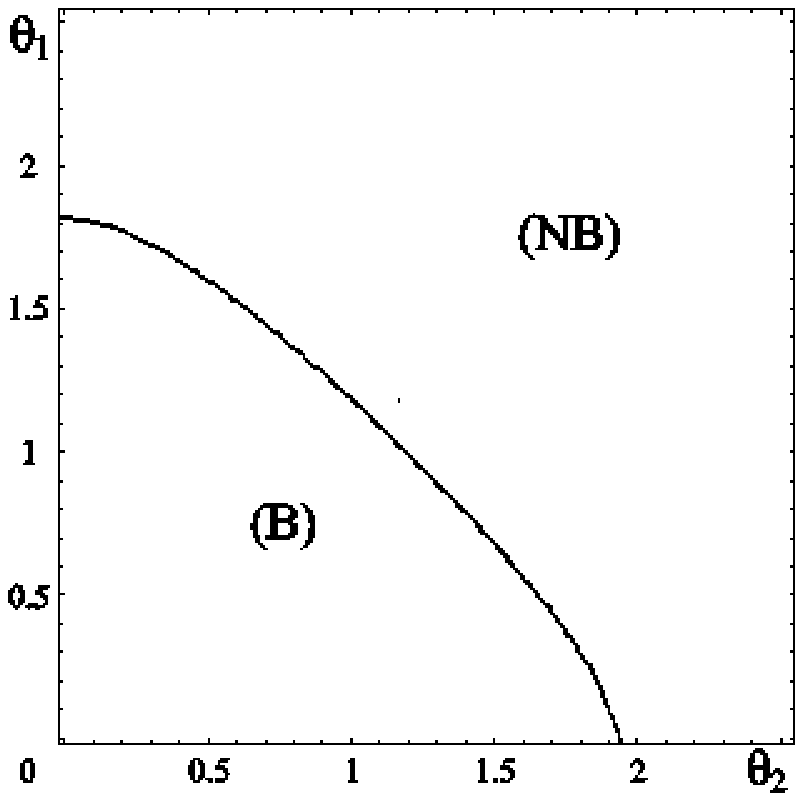} 
\caption{} 
\label{Fig. 4} 
\end{center} 
\end{figure}

\begin{figure}[htb]
\begin{center} 
\includegraphics{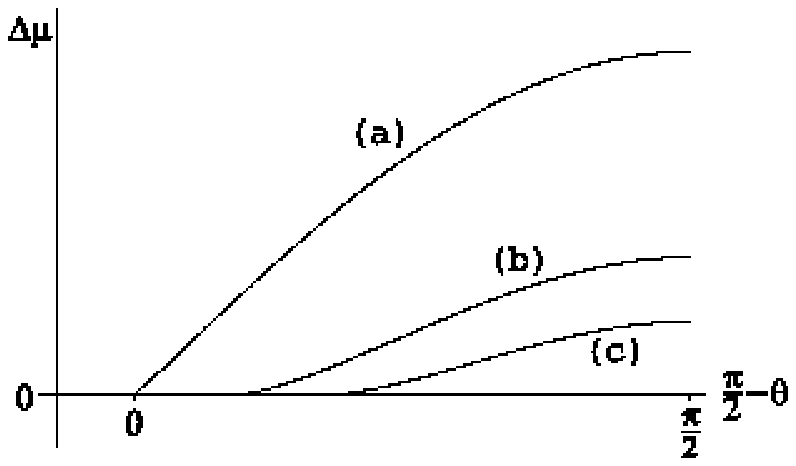} 
\caption{} 
\label{Fig. 5}
\end{center}
\end{figure}

In this note we are interested in the phase behavior of a fluid confined between two inert substrates forming the 
AFM-like geometry, see Fig.\ref{Fig. 1}. Our system is translationally invariant in $y$ direction.
The surface of the substrate $(1)$ is an infinite plane $z=0$ 
while the surface of the substrate $(2)$ at $z= \tan \varphi |x| +h$ is formed by two semi-infinite planes 
meeting at an angle $\pi - 2\,\varphi$; the line along which they meet is parallel to the first substrate 
surface. In the limit $\varphi =0, h\neq 0$ the system reduces to two  above mentioned wedge-like geometries, 
each with the opening angle $\varphi$.   \\ Depending on the values of the 
thermodynamic ($T, \mu$) and geometrical ($h, \varphi$) parameters the fluid confined in such AFM-like geometry  
can either display a configuration in which 
a liquid bridge connects the lower substrate with the AFM tip, or a configuration with no such liquid bridge 
present 
\cite{Bin}-\cite{Choe}. Our goal is to find the conditions under which the existence of such a liquid bridge in 
the above geometry is possible and to construct the relevant phase diagram. \\
Our approach is macroscopic and is based on minimization of the grand canonical free energy functional 
$\Delta \Omega [f]$ (per unit length in direction $y$ and relative to situation in which the space between the 
substrates is filled by vapor), 
where $f$ denotes the shape of the liquid-vapor interface. The minimum of $\Delta \Omega [f]$ is realized by the 
equilibrium 
interfacial shape $\overline f$. The functional  $\Delta \Omega [f]$ takes into account the relevant surface 
free energies as well as the appropriate bulk contribution in situation when the fluid thermodynamic state is off 
its bulk coexistence. It is parameterized by the corresponding substrate-liquid and substrate-vapor surface 
tension coefficients $\sigma_{il},\sigma_{ig}, i=1,2$, the liquid-vapor interfacial tension $\sigma_{lg}$, 
and by $\Delta \mu = \mu_{0} - \mu$. 
The surface tension coefficients depend on temperature and this dependence is transfered onto the relevant  
contact angles via the Young equation
\begin{eqnarray}
\label{Young}
\sigma_{ig} &=& \sigma_{il} + \sigma_{lg} \cos \theta_{i}, \quad i=1,2.  
\end{eqnarray} 
Note that he contact angles vanish at the corresponding wetting temperatures $T_{wi}$, $\theta_{i}(T_{wi}) = 0$. \\
The system under consideration is symmetric with respect to the $x=0$ plane, see Figs \ref{Fig. 1}, \ref{Fig. 2}
and this symmetry is reflected in the structure of the functional $\Delta \Omega [f]$. After using Eq.(\ref{Young}) 
it takes the following form  
\begin{eqnarray} 
\label{pot5}
\frac{\Delta \Omega [f]}{2\sigma_{lg}}  =  \int dz \left.\Bigg[ -\cos \theta_1 f(z)\delta(z) - 
\frac{\cos \theta_2 }{\sin \varphi}  + \sqrt{1+(\nabla f)^2}  \right. \nonumber \\   + 
\left.\frac{1}{\lambda} \Big( f(z) -a(z)\Big) \right.\Bigg]    \Theta \Big(f(z)\Big) 
\Theta\Big(f(z)-a(z)\Big)\Theta(z)   \\ + \cos \theta_2\frac{h}{\sin \varphi} - 
\frac{1}{\lambda} \frac{h^2}{ 2 \tan\varphi}   \nonumber  \quad,   
\end{eqnarray}  
where $a(z)=(z-h)/\tan \varphi$ describes the shape of substrate (2) surface, 
$\lambda = \frac{\sigma_{lg}}{\Delta \mu \Delta \rho}$, $\Delta \rho = \rho_{l}-\rho_{g}$ is the difference of 
the bulk densities, and $\Theta$ is the Heaviside function. Minimization of the functional $\Delta \Omega [f]$ 
with respect to $f(z)$ leads to the equation for the equilibrium profile $\overline f(z)$ 
\begin{eqnarray}  
\frac{d}{dz} \frac{{\overline f}'(z)}{\sqrt{1+{\overline f}'(z)^2}} &=& \frac{1}{\lambda} \label{rozn}  
\end{eqnarray} 
supplemented by two boundary conditions 
\begin{eqnarray} 
\label{warbrzeg1} 
\Big[ \frac{{\overline f}'}{\sqrt{1+{\overline f}'^2}}+\cos \theta_1 \Big] \delta(z) &=& 0   
\end{eqnarray} 
\begin{eqnarray}
\label{warbrzeg2}  
\Big[ \frac{1 + a'{\overline f}'}{\sqrt{1+ {\overline f}'^2}}-\frac{\cos \theta_2}{\sin \varphi} \Big] 
\delta({\overline f}-a)=0 \quad . 
\end{eqnarray} 
which can be rewritten in a geometrically transparent way 
\begin{eqnarray}   \left.\frac{df}{dz}\right|_{z=0} &=& - \frac{1}{\tan \theta_1} 
\label{brzeg1}\\  \left.\frac{df}{dz}\right|_{z=z_{2}} &=& \frac{1}{\tan (\theta_2 + \varphi)} \label{brzeg2} \quad. 
\end{eqnarray} 
The parameter $z_2$ is such that $f(z_{2}) = a(z_{2})$. Eq.(\ref{pot5}) contains the obvious constraint 
$f(z)>0$ for $0 \leq z \leq z_2$ on the allowed interfacial positions. Upon integrating Eq.(\ref{rozn}) one obtains 
the shape of the liquid-vapor interface in the form of the arc of a circle 
\begin{eqnarray}  
{\overline f}(z) = - \sqrt{\lambda^2 - (z - \lambda C_1)^2}+\lambda C_2 \label{fz} \ , 	 
\end{eqnarray} 
where $C_1 = \cos \theta_1$ and 
$ C_2 = [\cos(\theta_2 + \varphi)+\cos\theta_1 - h/\lambda]\cot \varphi + \sin(\theta_2 + \varphi)$ 
and $z_2 = \lambda \big[\cos \theta_1 + \cos (\theta_2 + \varphi) \big]$.  
After inserting the above function ${\overline f}(z)$ (\ref{fz}) into the functional in Eq.(\ref{pot5}) one obtains 
\begin{eqnarray}  
\frac{\Delta \Omega [{\overline f}]}{2\sigma_{lg}} = - 
\frac{(z_2-h)^2}{2 \lambda^2 \tan \varphi}  - \frac{z_2-h}{\lambda} \sin(\theta_2+\varphi)+\frac{A}{2} \quad , 
\end{eqnarray} 
where  parameter  
\begin{eqnarray} 
A(\theta_1,  \theta_2, \varphi) = \pi-\theta_1 - \theta_2 - \varphi + 
\frac{1}{2}[\sin(2  \theta_1) + \sin(2(\theta_2 + \varphi))]
\end{eqnarray} 
is non-negative for the range of angles we are interested in, i.e., $0 < \theta_1+\theta_2+\varphi<\pi$. It is 
straightforward 
to check that the condition for the existence of the bridge  $\Delta \Omega [f] \leq 0$ takes the following form 
\begin{eqnarray}
 \label{analkelv}
\Delta \mu \leq \frac{\cos \theta_1 + \cos (\theta_2 + \varphi)+
\tan \varphi \Big[\sin(\theta_2 + \varphi)-\sqrt{\sin^2(\theta_2 + \varphi) + 
\frac{A}{\tan \varphi}} \ \Big] }{\Delta \rho h  /\sigma_{lg}} \nonumber \\ 
\equiv {\cal F}(\theta_1, \theta_2, \varphi, h)  \quad. 
\end{eqnarray} 
The actual shape of the coexistence line 
$\mu_{AFM}(T) = \mu_{0}(T) -{\cal F}(\theta_1(T), \theta_2(T), \varphi, h)$  
between the phase with the bridge and without it depends on the way in which 
the contact angles $\theta_i$, $i=1,2$ depend on temperature. However, some general qualitative features of this 
coexistence  can be established independently from this temperature dependence. \\
First of all one notices that the width of the bridge, defined as - say - twice the minimum of 
$\overline f(z)$ \cite{Jang2},  is a decreasing function of the angle $\varphi$. In particular, in the limit 
$\varphi = 0$ at $h \neq 0$ our geometry reduces to that of a slit of width $h$, 
and the Eq.\ref{analkelv} reduces to the Kelvin law 
\begin{eqnarray}
\label{kelv}
\mu_{h}(T) = \mu_{0}(T) - {\cal F}(\theta_1, \theta_2, 0, h) = 
\mu_{0}(T) - \frac{[\cos \theta_2 + \cos \theta_1] \sigma_{lg}}{\Delta \rho} \,\frac{1}{h}\quad,
\end{eqnarray}
which says that for $\Delta \mu \leq {\cal F}(\theta_1, \theta_2, 0, h)$ "an infinitely thick liquid bridge" 
exists in the slit. Another limiting case is obtained for $h=0, \varphi \neq 0$ and corresponds to two wedges 
touching each other. For simplicity we assume chemically identical substrates, i.e., 
$\theta_{1} = \theta_{2} = \theta$. 
To analyze this special case we first define the parameter $\ell_{h}$ as in  Fig.(\ref{Fig. 3}).  
A straightforward calculation leads to    
\begin{eqnarray} 
\label{grubosc_h}
\ell_h  =  \frac{\sigma_{l g}}{\Delta \mu \Delta \rho} \cdot \frac{\cos \theta - 
\sin \frac{\varphi}{2}}{\sin \frac{\varphi}{2}} - \frac{h}{\tan \varphi \cos \frac{\varphi}{2}}	\quad.  
\end{eqnarray} 
Upon approaching the coexistence curve $\mu \to \mu_0$ the parameter $\ell_h$ tends to infinity. In particular, 
for $h=0$ the above formula takes the form  identical to the one derived in  \cite{Rejmer}. \\ We also note that the 
rhs of Eq.(\ref{analkelv}) vanishes for the 
values of the contact angles $\theta_{1}, \theta_{2}$ such that the condition for the filling transition of the wedge 
at temperature 
$T_{f}(\varphi)$ ($[\theta_{1}(T_{f}(\varphi)) + \theta_{2}(T_{f}(\varphi)]+\varphi = \pi$) is satisfied, 
see \cite{Rejmer,Jakubczyk}.  \\
The relevant phase diagrams are displayed in Figs \ref{Fig. 4}-\ref{Fig. 5}. 
A particularly transparent form of the phase diagram is presented on Fig.\ref{Fig. 5}.   
We note that upon increasing $\mu$ at fixed temperature the system first undergoes a discontinuous transformation at 
which a liquid bridge is formed. This takes place at 
$\mu = \mu_{AFM}(T) = \mu_{0}(T) - {\cal F}(\theta_1, \theta_2, \varphi, h)$, see  Eq.(\ref{analkelv}). Upon further 
increase of 
$\mu$ one observes a continuous growth of the parameter $\ell_h$ in accordance with the law obtained for 
the complete filling in the wedge-like geometries. Thus the phase diagram for the fluid confined in the 
AFM-like geometry in Fig. \ref{Fig. 5} combines both the features typical for the capillary condensation 
and the filling of the wedge. Once the bridge has been formed it's thickness growth upon approaching the 
bulk coexistence in accordance with the law obeyed by the case of complete filling in the corresponding wedge.  
\\

\begin{figure}[htb]
\begin{center} 
\includegraphics{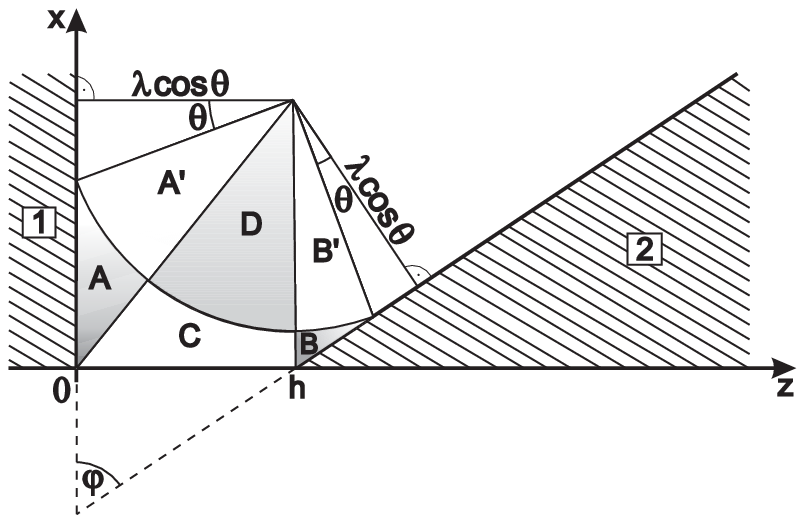} 
\caption{}
\label{Fig. 6}
\end{center}
\end{figure}

Finally, we note that part of the above analysis can be formulated in purely geometric terms. 
Eqs(\ref{brzeg1}-\ref{fz}) determine the contact angles and the shape of the vapor-liquid interface as an arc of
a circle with radius $\lambda$. In the case of  identical substrates, i.e., equal contact angles 
$\theta_1 = \theta_2 = \theta \leq \pi/2-\varphi $ the expression for the   functional $\Delta \Omega$ can be written 
in the following way 
\begin{eqnarray}
 \frac{\Delta \Omega}{2 \Delta \mu \Delta \rho} = ( C + 2D ) - ( A + B) \equiv S(h) \quad,
\end{eqnarray} 
where the symbols $A, B, C, D$ denote the areas of the corresponding regions in Fig. \ref{Fig. 6}.  
The shift of the position of the AFM tip at fixed contact angles (i.e., temperature) and the radius 
$\lambda$
(i.e., chemical potential) one be simulated on Fig. \ref{Fig. 6} by changing the position of the 
$x=0$ axis. In particular, for $h=0$ the areas $D$ and $C$ vanish, hence $S(0) < 0$, 
and the existence of the liquid bridge is favorable. On the other hand, for 
$h = h_t = \lambda \tan \varphi[\frac{\cos \theta}{\tan \varphi /2}-1]$    
(the vapor-liquid interface being tangent to the $x=0$ axis) the geometry dictates $S(h_t) > 0$, and the 
liquid bridge ceases to exist. \\
Thus the function $S(h)$ has opposite sign in these two cases. Since it depends on $h$ in continuous way,  
there exists $0 < h_{AFM} < h_t$ such that $S(h_{AFM})=0$; at $h=h_{AFM}$ the 
phase coexistence takes place for given temperature and chemical potential. \\

To summarize, we showed that the capillary condensation in a slit and the wedge filling phenomena can be treated as 
special cases of bridge formation in the AFM-like geometry. In particular, we obtained 
the shape of the liquid bridge, the function describing the shift of the coexistence curve 
${\cal F}(\theta_1, \theta_2, \varphi, h)$, and the corresponding  phase diagrams.

\newpage
\begin{center}
\bf{Figure captions}
\end{center}

\begin{itemize}
\item[Fig. 1]  A bridge of the liquid 
phase ($l$) surrounded by the gas phase ($g$) in AFM-like geometry. The system is translationally invariant 
in  direction $y$; the fluid is bounded from below by the surface $z = 0$  of the lower substrate (1) and from  
above by the surface $z= \tan \varphi |x| +h$ of the upper substrate (2), where $h$ denotes the distance 
between the tip and the lower substrates. \\
\item[Fig. 2] One half of the system from Fig.\ref{Fig. 1}. The interface $x=f(z)$ in the form of an arc of a circle 
separates the liquid ($l$) and vapor ($g$) phases. The contact angles $\theta_1$ and $\theta_2$ are displayed and 
the parameter $z_2$ is such that $f(z_2)=(z_2-h) \cot \varphi$. \\
\item[Fig. 3] The liquid-vapor interface has the form of an arc of a circle. 
In the limit $ h \to 0$ the parameter $\ell_h$ corresponds to the height of the liquid-vapor interface at 
the center of the wedge. \\
\item[Fig. 4] The phase diagram displayed in variables $\theta_2$ and $\theta_1$ for the specific choice of parameters 
$h/\lambda =0.1 $ and $\varphi = \pi/6$. 
The coexistence curve separates the phase with the bridge (B) and the phase without it (NB). \\
\item[Fig. 5] The phase diagram in the case of identical substrates ($\theta_1=\theta_2 = \theta$) displayed in 
variables $\frac{\pi}{2}-\theta$  and $\Delta \mu$ (measured in $\frac{\sigma_{lg}}{\Delta \rho h}$ units) for three choices 
of system parameters: (a) $\varphi = 0$, (b) $\varphi = \pi/6$, (c) $\varphi = \pi/3$. 
 \\
\item[Fig. 6] The liquid bridge in case of identical substrates ($\cos \theta_1 = \cos \theta_2 = \theta$). The 
volume of the adsorbed liquid is equal to the sum of the areas $A+B+C$, while 
$\lambda f(0)=2(A+A')$, and    
$\lambda f(z_2) / \cos \theta =2(B+B')$. The product of $\lambda$ and the length of the gas-liquid interface 
is equal to $2(A'+D+B')$.
\end{itemize}

\newpage
\vfill
\vfill
\vfill

\newpage
\vfill
\vfill
\vfill

\newpage
\vfill

\newpage 

\end{document}